# Publisher References in Bibliographic Entity Descriptions

**Jim Hahn**
University of Pennsylvania
Penn Libraries, USA
jimhahn@upenn.edu

**ABSTRACT**

This paper describes a method for improved access to publisher references in linked data RDF editors using data mining techniques and a large set of library metadata encoded in the MARC21 standard. The corpus is comprised of clustered sets of publishers and publisher locations from the library MARC21 records found in the POD Data Lake, an Ivy+ Library Consortium metadata sharing initiative. The POD Data Lake contains seventy million MARC21 records, forty million of which are unique. The discovery of publisher entity sets described forms the basis for the streamlined description of BIBFRAME Instance entities. This study resulted in two major outputs: 1) A prediction database and 2) sets of publisher location and name association rules. The association rules are the basis of a prototype autosuggestion feature of BIBFRAME Instance entity description properties designed specifically to support the autopopulation of publisher entities in linked data RDF editors.

**INTRODUCTION**

Library practitioners are progressing from cataloging practices of library metadata encoded in MARC21 to linked data encoded in BIBFRAME RDF (Library of Congress, 2016). The LD4 community's recent advances and the Library of Congress's work have allowed library descriptions in linked data to be created with native RDF. A major contribution among the LD4 grants and community include the open source linked data RDF editor, Sinopia (Nelson, 2019). The Library of Congress is engaged in a year-long process to integrate BIBFRAME entity descriptions into their systems fully (Library of Congress, 2021). While both traditional metadata and new RDFbased metadata will exist side by side for some time, early adopters of BIBFRAME have a growing need to streamline the creation of linked data entity descriptions.

**Bibliographic Entities as Sets**
In a position paper on "Bibliographic Entities and their Uses," Svenonius (2018) posited the set theoretical view that "Bibliographic entities are sets of items…" and argued that "… one item is taken to be emblematic of each item in the set. The attributes of the item are, by way of shorthand, regarded as attributes of the bibliographic entity to which the item belongs" (p. 715). Using a set theoretical frame for bibliographic entities is the departure point for this research. Properties of bibliographic entity descriptions may be identified using frequent pattern data mining algorithms over targeted sets of associated legacy metadata. Not all sets or MARC21 tag associations will be of interest for data mining. Associations in the legacy cataloging description vary in their application and use – in one sample it was found that a small percentage (less than 5%) of MARC21 tags accounted for almost 80% of use (Moen & Benardino, 2003).

Semi-automated bibliographic entity identification using data mining outputs may streamline BIBFRAME cataloging to allow rapid identification of the authority properties required in BIBFRAME descriptions. In this paper, the focus is on the BIBFRAME Instance description. Selecting and referencing linked open references is time consuming for the linked data cataloger, with many potential open linked data references to incorporate. The BIBFRAME data model's core hierarchy is comprised of BIBFRAME Works and Instance entity descriptions. A BIBFRAME Work entity description is "…the highest level of abstraction," and, further, "…reflects the conceptual essence of the cataloged resource: authors, languages, and what it is about (subjects)" (Library of Congress, 2016, p. 1). The BIBFRAME Instance entity description is concerned with "… one or more individual, material embodiments, for example, a particular published form" (Library of Congress, 2016, p. 1). In practice, Work and Instance BIBFRAME entity descriptions reference linked open data on the web, such as the Library of Congress subject files at id.loc.gov. The resources at the id.loc.gov authority service needed most commonly include the Library of Congress Subject Headings and the Name Authority File.

**DATA MINING METHODS**

Data mining methods in this project can be understood as an extract, transform, and load (ETL) process. The extraction of associative publisher name and place from legacy Ivy+ metadata creates comma separated values (CSV) from MARC field data. To address the varieties of metadata quality, the CSV are then loaded into OpenRefine, a data cleaning tool, to cluster publisher name and place. OCLC research's previous work, which experimented with the development of a publisher name authority file from existing metadata, inspired clustering over a publisher's name (Connaway, Silipigni, & Dickey, 2011). The clustered associative publisher name and place data are then loaded into a data frame using a PySpark Jupyter Notebook, which is configured thereafter to make use of SparkML functionality, specifically the Fp-growth algorithm (Han, Pei, & Yin, 2000).

**Extract: Obtaining Data**

The POD Data Lake completed the proof-of-concept work for the Minimum Viable Product (MVP) in 2020 and makes Ivy+ Library metadata available to its membership. Library metadata encoded in the MARC21 standard vary by library. The table below delineates the corpus accessed from the data pool for this project. Note that not all Ivy+ Libraries are represented in the corpus. To complement the corpus of records available, enriched PCC Data Pool records (Samples & Bigelow, 2020) and Library of Congress open metadata were added.

| Library/Dataset Source | MARC21 Records | Sets {publisher name, publisher place} {MARC 260$b, MARC 260$a} |
|---|---|---|
| Columbia University – 2020 | 476970 | 460464 |
| Dartmouth University – 2020 | 1152503 | 1012040 |
| Library of Congress Open Metadata Set – 2016 | 2670462 | 2601806 |
| PCC Data Pool Export 1 – 2020 | 4263628 | 2593909 |
| University of Pennsylvania – 2020 | 5109538 | 4373619 |
| Brown University – 2020 | 5490307 | 4454205 |
| Duke University – 2020 | 6704722 | 5209400 |
| University of Chicago – 2020 | 7648280 | 6512061 |
| Stanford University – 2020 | 8258948 | 6288443 |

**Table 1. Data Corpus in this Research is Comprised of a Selection of Early Access POD Data Supplemented with LC and PCC Data Pool Metadata. Dates Indicate When Data Were Published Originally**

**Transform: Clustering Data**

Several filtering and clustering steps were conducted to make the data mining process effective. Specifically, transformation is aimed to eliminate any potential "noise" in the dataset. Datasets were first filtered for sets that contained the publisher's name and location. Then, after filtering, various stop characters were also discarded or excluded if they contained characters that do not represent words accurately. For example, these libraries' historical cataloging practice used brackets to indicate uncertainty or assumed location or publishers. Bracket data were discarded in favor of more certain strings of text. Strings of the remaining text in the legacy MARC21 are known to be entered as uncontrolled names. The next step in the process therefore, used OpenRefine text clustering (https://openrefine.org/).

Previous work has found that text clustering of publisher names is effective (Connaway, Silipigni, & Dickey, 2011). The OpenRefine implementation used a common algorithm method of "fingerprint key collision," that is known to be both quick and results in few false positives. According to the OpenRefine release notes, many of its text clustering algorithms available for fingerprinting were developed initially in the SIMILIE project at MIT (Butler et al., 2004). In turn, these were influenced by several seminal overview works on pattern matching and string processing in (Hull, 1996; Navarro, 2001).

As an output of the clustering techniques in OpenRefine, 2.4 million (2,460,674) unique publisher names were found in the corpus of the over 40 million records. Uniqueness is defined as one and only one occurrence of an identified string of either publisher name or city after the filtering and clustering aforementioned. After clustering the city names and de-duplicating the initial set of 40 million records, the same filtering and clustering technique found that a smaller set of 314,244 cities unique cities was represented in the dataset.

*Transform Inputs*

MARC21 Records Retrieved for Data Mining: 41,775,358
MARC21 Records with Publisher Name/Publisher Place Sets: 33,505,947

*Transform Outputs*

Filtered and Clustered Training Corpus: 24.5 million non-unique pairs
- Unique Publisher Names: 2.4 million (2,460,674 clustered names)
- Unique Publisher Cities: 314,244

**Load: Spark Machine Learning**

The Fp-growth algorithm can be called from within Spark machine learning software. Spark is a software tool used commonly to compute clusters in big data processing and scales very well to large datasets. Data mining researchers introduced Fp-growth over twenty years ago and its techniques still help identify interesting patterns in datasets (Han et al., 2000). A refinement to parallelize the algorithm for automated query suggestion was presented in a publication in the ACM Recommender Systems conference proceedings (Li, et al., 2008). Settings for Fp-growth support and confidence can be configured within the Juypter notebook. As an experimental first step in data mining of all association rules, minimum support is set at 0.00001 and the minimum confidence set as 0.6, shown in Figure 1.

```
Spark FP Growth using code samples from Spark docs for python parallel fp growth alogorithm:
https://spark.apache.org/docs/latest/ml-frequent-pattern-mining.html

import os
import pyspark
import pyspark.ml.fpm
from pyspark.context import SparkContext
from pyspark.sql.session import SparkSession
from pyspark.sql.functions import countDistinct
from pyspark.ml.fpm import FPGrowth
sc = SparkContext('local')
spark = SparkSession(sc)

#read json file into dataframe
df = spark.read.json("./work/*.json")

#df.show()
df.count()

24590867

#set model support and confidence
fpGrowth = FPGrowth(itemsCol="items", minSupport=0.00001, minConfidence=0.6)

model = fpGrowth.fit(df)
```

**Figure 1. Configuration of the FPGrowth Algorithm in the Juypter PySpark Notebook. Inspired by code from: https://spark.apache.org/docs/latest/ml-frequent-pattern-mining.html**

**RESULTS**

The results from data mining are comprised of rule and prediction databases. Sample data from each of these databases are shown in Tables 2 and 3. The full association rules database table can be viewed here: http://ow.ly/plC550Erq0S

**Association Rules**

The association rules database contains 7,937 rows of data grouped into 5 columns comprised of: antecedent, a text column with 7,937 unique values; and consequent, a text column with 1563 unique values. The most frequent values in the consequent column are London (899), New York (836), Paris (441), Washington D.C. (239), and Moskva (204). There are three numeric columns in the association rule table: confidence, lift, and support. Confidence and support metrics act to guide evaluation of fp-growth data mining and were introduced in association rule algorithms (Agrawal & Srikant 1994). Lift is a metric that builds upon confidence and support computations (Bayardo & Agrawal, 1999).

| **Antecedent** | **Consequent** | **Confidence** | **Lift** | **Support** |
|---|---|---|---|---|
| American Library Association | Chicago | 0.954449986873195 | 111.876299789079 | 0.000295678879479931 |

**Table 2. Schema and Sample Row from Rule Database Prediction Database**

This prediction database is made up of two text columns: an item set row, and the prediction set row. There are a total of 1,599,051 associative rows from which an RDF linked data editor may derive property autosuggestions.

| **Items** | **Prediction** |
|---|---|
| [Chicago, Law Student Division American Bar Association] | [University of Chicago Press] |

**Table 3. Schema and Sample Row from Prediction Database**

**CONCLUSION AND FUTURE WORK**

In this project, data were clustered syntactically based upon rules for lexical parsing. If the uniform sets that are clustered syntactically now can be reconciled further with linked data entities for which their attributes are members, it follows that semantic clustering can be achieved. The outcome of semantic clustering would streamline cataloging further by obviating the need for many authority entity data to be identified in RDF-linked data editors.

In this paper, the set theoretical approach based on set definitions made use of Svenonius's (2018) work as a departure point to discover the way a corpus of existing legacy metadata attribute sets may ontologically–by set existence–describe publisher entities. Further semantic clustering may discover entities that exist ontologically in linked open data form. Linked open data on the web is comprised of dereferenceable URIs whose properties might be found in patterns of MARC21 data.

To streamline BIBFRAME metadata creation further for linked data catalogers, future work includes the development of association rules based upon associations from Agent Entity to a Publisher Entity autosuggestion, and the Author Entity to Subject Entities for autosuggestion in a linked data RDF editor. Furthermore, as the corpus of linked data Work entity description sets grow, there is also the future possibility to perform semi-automated predictions for related Works or Super Works by attribute set membership.

An anticipated outcome of this research, poised for integration into practice, will make possible semi-automated RDF editors whereby linked data creation becomes a much less resource consuming activity than legacy processes of metadata creation and maintenance.